# Simulating urban expansion in the parcel level for all Chinese cities

Ying Long
Beijing Institute of City Planning, Beijing, China
University of Cambridge, UK

Kang Wu
Capital University of Economics and Business, China

Qizhi Mao
Tsinghua University, China

**Abstract**: Large-scale models are generally associated with big modelling units in space, like counties or super grids (several to dozens $km^2$). Few applied urban models can pursue large-scale extent with fine-level units simultaneously due to data availability and computation load. The framework of automatic identification and characterization parcels developed by Long and Liu (2013) makes such an ideal model possible by establishing existing urban parcels using road networks and points of interest for a super large area (like a country or a continent). In this study, a mega-vector-parcels cellular automata model (MVP-CA) is developed for simulating urban expansion in the parcel level for all 654 Chinese cities. Existing urban parcels in 2012, for initiating MVP-CA, are generated using multi-levelled road networks and ubiquitous points of interest, followed by simulating parcel-based urban expansion of all cities during 2012-2017. Reflecting national spatial development strategies discussed extensively by academics and decision makers, the baseline scenario and other two simulated urban expansion scenarios have been tested and compared horizontally. As the first fine-scale urban expansion model from the national scope, its academic contributions, practical applications, and potential biases are discussed in this paper as well.

**Key words**: urban expansion simulation; vector cellular automata; applied urban modelling; land parcel; MVP-CA

**Acknowledgments:** We acknowledge the financial support of the National Natural Science Foundation of China (No.51078213 and 51311120081) and China Scholarship Council (No.201209110066). Our thanks also go to Dr Dong Li for providing the ordinary survey map of China. We would release the simulation results online on acceptance of this paper.

# 1 Introduction

This study develops a vector cellular automata model for simulating urban expansion in the parcel level for all Chinese cities within the context that most of urban expansion models made a balance between global spatial scale and local details. Our model makes some progress on simulating both for a large scale (the whole China) and with fine-level spatial units (the parcel level[1]).

Since the Reform and Open Policy China's rapid urbanization has been exerting a far-reaching influence on the evolution of world's civilisation and human society (Montgomery, 2008; Liu et al, 2012). China's total urban population has increased from 172 million to 712 million between 1978 and 2012, and the number of reported cities at all levels has reached 656 in 2012 (tripled that in 1978) (National Bureau of Statistics, 2013). Meantime, China is also experiencing an unprecedented active stage of urban expansion. By the end of 2012, urban land in China has reached 46,751 $km^2$, increased 624.7% compared to 1983. Consequently, Chinese cities are bearing increasingly heavy burden on natural resources and ecological environment (Liu and Diamond, 2005). Under such a social-economic background, urban expansion of Chinese cities has attracted extensive attention internationally and locally on identifying urban spatial morphology and growth boundaries (Long et al, 2013), monitoring temporal-spatial process and pattern (Seto et al, 2002; Tian et al, 2005; Liu et al, 2010; Kuang, 2012), detecting driving forces and mechanism (Seto et al, 2003; Deng et al, 2010), simulating temporal-spatial process (Kuang, 2011), analyzing future scenarios (Kuang, 2011), as well as assessing ecological and environmental impacts. (Gong et al, 2009). In total, spanning different spatial scales, many efforts have been made in China's urban expansion research but some limitations continue to exist obviously. Firstly, much attention has been paid in describing the temporal-spatial patterns, especially in the national and regional scale, while temporal-spatial simulation and future scenarios analysis were relatively rare. Some researches explore the simulation methods and analyze the impending scenarios generally based on a micro-level or a middle-level scale but fail to meet the needs of urban dynamic spatial modelling in large scales.

Better understanding future urban expansion is essential for policy design and its effectiveness predication. Currently, there are several mature approaches developed for simulating urban expansion dynamically, which at least include cellular automata (CA), agent-based modelling (ABM), CLUE or CLUE-S (Conversion of land use and its effects), artificial neural network (ANN), and system dynamics (SD). Incorporated with other models or methods like SD, ABM and ANN in some cases, CA models have become well-established tools for modelling urban expansion thanks to their ability to simulate dynamic spatial processes from a bottom-up perspective (Santé et al, 2010). They have been applied in several studies of Chinese city regions like Beijing (Long et al, 2009), Northern China (He et al, 2006), Guangzhou and Pearl River delta (Li and Liu, 2008; Liu et al, 2008), and Beijing-Tianjin-Tangshan metropolitan area (Kuang, 2011), For these grid CA models mentioned above, geographic space is typically represented as a grid of regular cells (ranging from 50 m to 1km in Chinese cases) and the neighbourhood is defined as a assembly of adjacent cells.

---

[1] Parcels in China correspond to "blocks" in western countries like the USA.

Recent studies have demonstrated that such raster-based CA models are sensitive to the modifiable units used in the models. For instance, Chen and Mynett (2003) investigated the effects of cell size and neighbourhood configuration in a prey-predator CA model and observed that they affected both the resulting spatial patterns and the system stability. Jantz and Goetz (2005) examined the results of SLEUTH model in response to different cell sizes and indicated that the cell size at which the land use data were represented could impact the quantification of land use patterns and descriptive power of the model Therefore, grid CA is proven to be an efficient method to simulate dynamic spatial processes, but it's necessary to improve the method from the traditional raster-based perspective to other new ones.

Some researchers started to use vector, or irregular, cells rather than the traditional regular ones in CA models to avoid the questions mentioned above. In recent years, the studies of vector-based CA have gained many focuses importantly from academics. O'Sullivan (2001) combined CA and graph theory to generate sets of neighbourhood-scale irregular cells. Moreover, irregular cell usually was employed to represent entities in real world. For instance, Torrens and Benenson (2005) proposed the geographic automata system (GAS) combining characteristics of both CA and multi-agent models with the aim for incorporating irregular vector objects as automata to represent real-world entities such as roads, buildings and parks. Stevens and Dragicevic (2007) developed *iCity*, in which an urban area is partitioned into discrete land use units based on cadastral information and represented as a collection of polygons. Shen and Kawakami (2008) developed a geosimulation model using the vector-based CA to visualise land use patterns in urban partitions. Pinto and Antunes (2010) developed an irregular CA based on census blocks to determine the land use demand under the consideration of dynamics of population and employment densities over time. In the entity-based CA model presented by Menard (2008) and Moreno et al (2009), the shape and size of each object can also change and a dynamic neighbourhood could be semantically implemented. In short, irregular polygons are more representative to the real world, e.g. parcels and blocks, while raster cells are not directly corresponding to real geographical entities.

Existing large-scale urban expansion models seldom use vector polygons, especially small-scale parcels, as cells in CA. Simulating urban expansion for a large urban extent in a fine scale is promising as follows. (1) Parcels, as clear behavioural units, would be more appealing to local decision makers and citizens, since each parcel rather a regular cell has a fixed boundary bounded with local images and knowledge; (2) Land use regulations could be distributed to parcels directly, and each city would have access to the simulation results. This would benefit those cities with no financial or intelligent stock supervising or being aware of future developments through our model.; (3) The simulation results could be compared in the city level in the fine scale so that some intra-city phenomenon could be observed. (4) Such model enables integrating spatial interaction analysis (flows and networks) for future.

However, the fine-scale urban expansion model for a large area is rare in academic research to the best of our knowledge, generally due to (1) fine-scale parcel/block data availability, (2) computation limitation. Data limitation is more serious in China than other countries, for example, the best available parcel map for China's capital Beijing – supposedly one of the most technologically advanced and rapidly developing cities in the erstwhile Third World – was dated in 2010 (Beijing Institute of City Planning 2010). In addition, collecting parcel data in medium and small sized cities in China is constrained by poorly developed digital infrastructures. Not only hard infrastructure but also soft institutions hamper Chinese urban planners' access to parcel maps. For instance, our interviews with 57 planning professionals reveal that access to existing parcel maps held by local planning bureaus/institutes is extremely controlled, as parcel maps are tagged as confidential within the current Chinese planning institutions. In summary, parcel data for the developing world is oftentimes outdated and limited in geographical scopes. This condition, to some degree, has obstructed the progress of urban expansion model for a large area in fine-scale in developing countries in general and in China in particular. Overcoming such "data desert" scenario seems to be the first priority for fine-scale urban simulation in developing countries.

In this paper, a mega-vector-parcels cellular automata model (MVP-CA) is developed for simulating urban expansion in the parcel level for all 654 Chinese cities. To secure the precondition, we would use our established framework for generating parcels and identifying urban parcels from all generated parcels for preparing the base map for future fine-scale urban expansion simulation (Long and Liu, 2013). This technique has been applied for using roads in OpenStreetMap to partition regional space, and inferring urban parcels using well-classified POIs provided by online map providers. The full process is automatic and could be easily adapted to annually update existing parcel map supported with upgraded OSM and POIs. We would update the OSM roads with the detailed road network in ordnance survey to generate urban parcels in all cities in association with POIs in China. In this regard, the problem of the base map for the urban expansion simulation for China is then solved. Also, more effort is paid on the computation load detailed in the method section. This paper is structured as follows. Section 2 describes the datasets used in this paper. The methods and their results are introduced in Section 3 in detail. We would compare the generated parcels together with their attributes with existing data source. Section 4 and 5 discuss the results and make concluding remarks of this research, respectively. Finally, section 6 draws a conclusion.

## 2 Data

### *2.1 Administrative boundaries of Chinese cities*

A total of 654 cities in China are analysed in this study(Figure 2)[2], ranging across five administrative levels: namely municipalities directly under the Central Government (MD, 4 cities), sub-provincial cities (SPC, 15), other provincial capital cities (OPCC, 17), prefecture-level cities (PLC, 250), and county-level cities (CLC, 368) (Ministry of Housing and Urban Development, MOHURD, 2013; see Ma, 2005 for more details about the Chinese administrative system). As a city proper in China contains both rural and urban land uses, our analytical scope is narrowed down onto legally defined urban land within city proper.

---

[2] Sansha in Hainan and Beitun in Xinjiang appearing in MOHURD (2013) were not included due to spatial data availability. Taiwan was not included in all analysis and results in this paper.

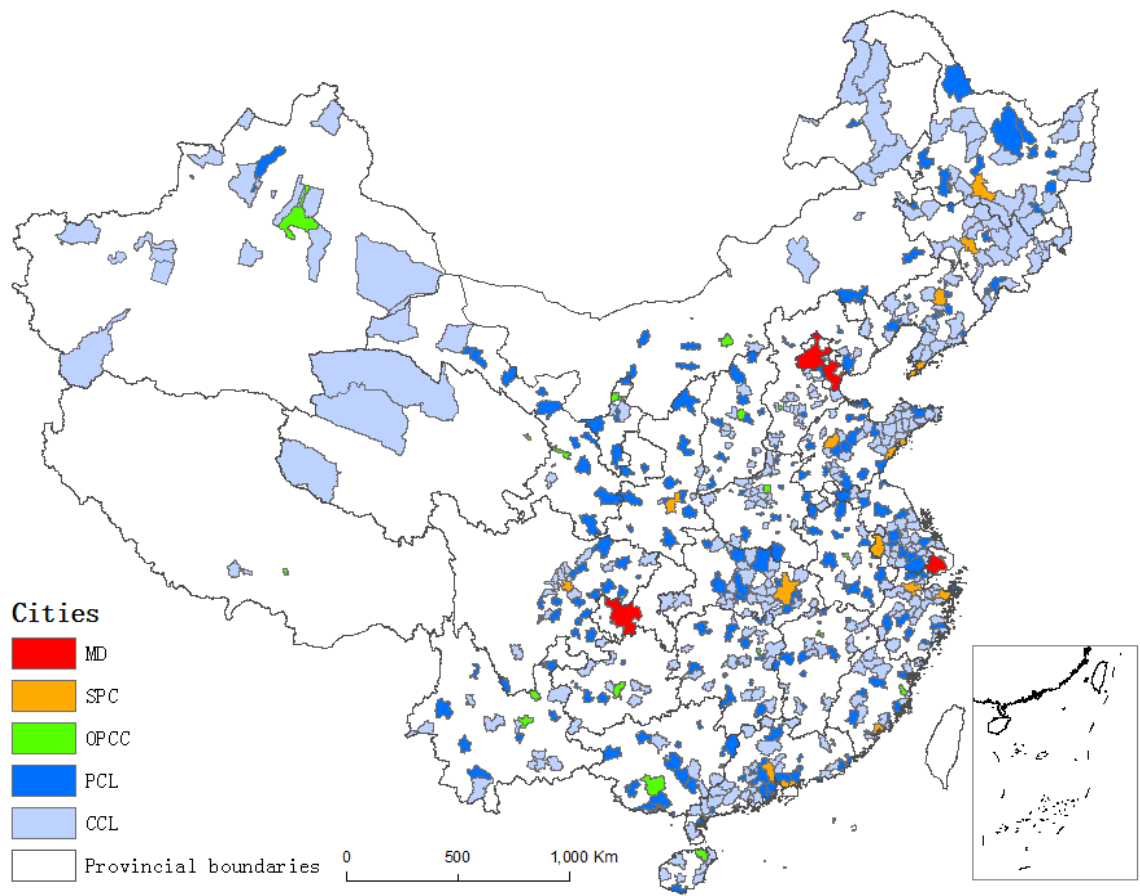

**Figure 1 Administrative boundary of Chinese cities**

## *2.2 Urban land area in 2007 and 2012*

In addition to administrative boundaries, urban land area for each Chinese city is required in 2012 to calibrate the urban expansion rate for all cities. To achieve this aim, information about total urban lands of individual cities from 2007 to 2012 is collected from MOHURD (2013). As the result of Chinese urbanisation, adjustment of administrative divisions occurs frequently every year, making the city number in each level changes correspondently. For consistency, some city boundaries are merged and revised according to the latest administrative districts and city inventories in 2012 to ensure every single city comparable on the time dimension. According to the statistics, total urban land area of 654 cites in China reached 36,352 and 46,744 km$^2$ in 2007 and 2012 respectively. The average growth rates of 654 cities raised to 4.5% in past five years. Furthermore, the growth rate of every single city during 2007-2012 was estimated for which setting the business-as-usual urban expansion scenario.

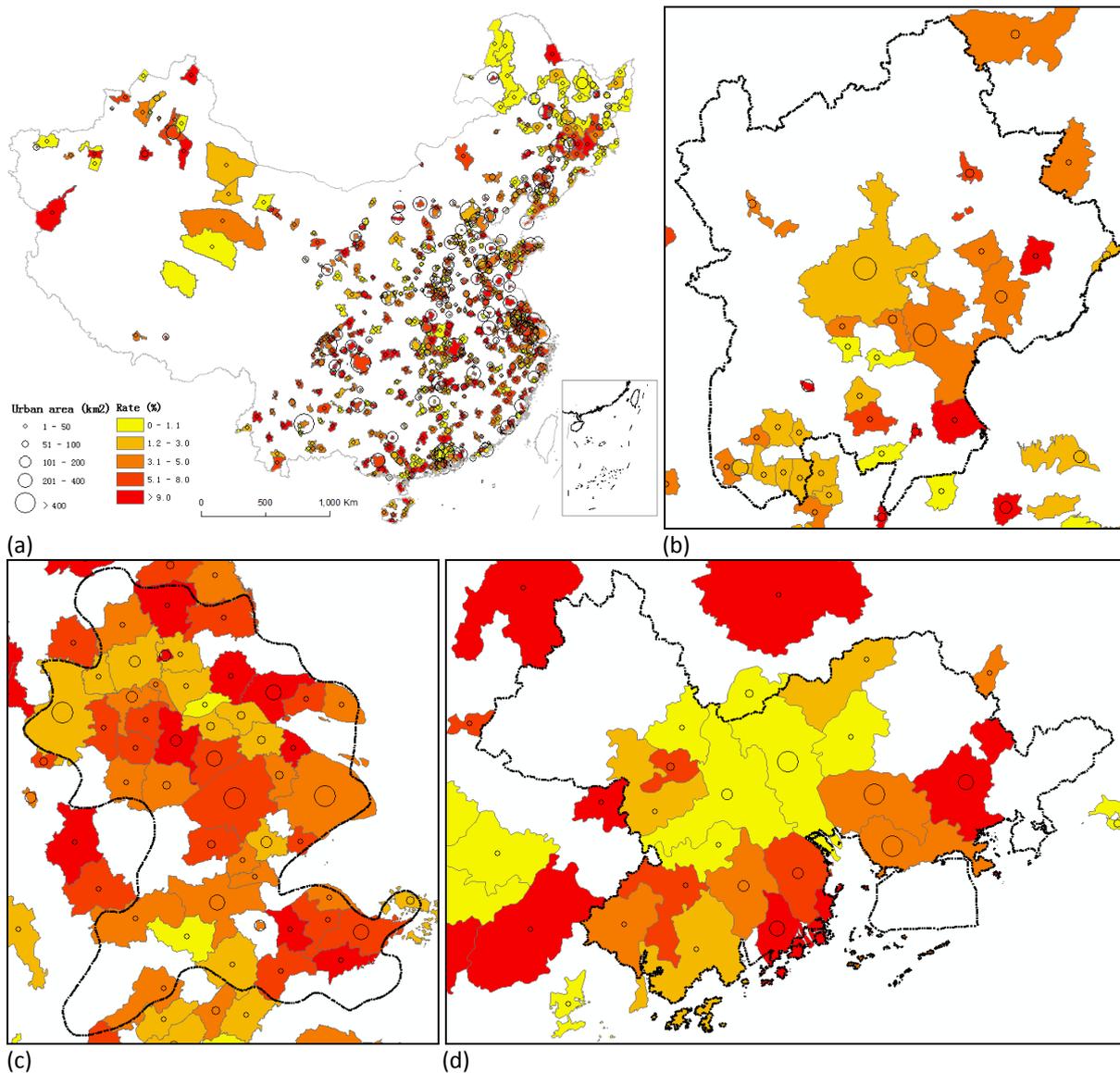

**Figure 2 Urban land area in 2012 and the urban expansion rate during 2007-2012 (a) the whole China, (b) Beijing-Tianjin-Hebei (BTH), (c) Yangtze River Delta (YRD), (d) Pearl River Delta (PRD)**

## *2.3 The ordnance survey roads and POIs in 2011*

Two datasets are used for generating the urban parcels of all Chinese cities. The first one is the ordnance survey map of China in 2011 with detailed road networks obtained from a local road navigation firm based in Beijing. Almost all detailed road networks in various levels including streets and regional roads were included in this data according to the examination via comparing with Google Maps and Baidu Map (a main online map service provider, also the dominating online search engine in China, http://map.baidu.com). Total road length was 2,623,867 km for 6,026,326 segments[3] (Figure 3).

---

[3] We would use the term "road" for all terms of transportation network elements, like streets and highways.

A total of 5,281,382 POIs are gathered and geo-coded by business cataloguing websites. The initial twenty POI types are aggregated into eight more general assemblies: Commercial sites account for most POIs, followed by business establishments, transportation facilities, and government buildings. POIs labelled as "others" are used in estimating land use density, but removed from land use mix analysis as this type of POIs with mixed information are not well organized and classified according to our review. One the other hand, the data quality is secured through manual checking randomly selected PODs. More importantly, this empirical framework is extensible in the sense that POI counts can be replaced by other human activity measurements, ranging from the more conventional land use cover derived from remote sensing images to ubiquitously available online check-in service data (e.g., Foursquare) in the background of web 2.0.

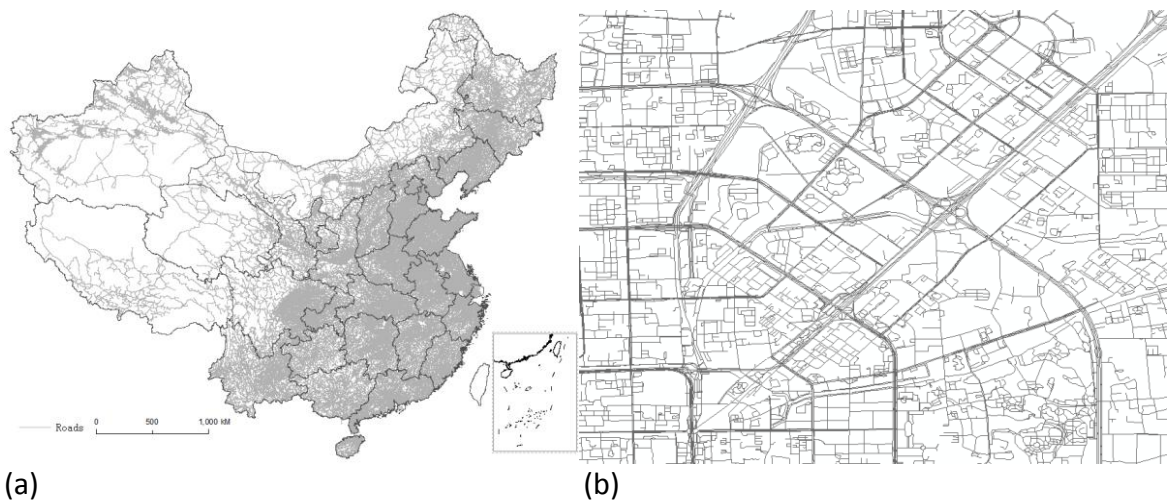

(a)                                          (b)

**Figure 3 Ordnance roads of China in 2011. a: the whole China; b, a part of the central city of Beijing**

## *2.4 Other data*

The distance to city centres for all parcels is considered in the proposed MVP-CA model. The administrative centre, generally also the city centre, of each city has been manually prepared as a point layer in GIS. In addition, two layers regarding with natural limitation for construction are selected as the exclusive development area in the MVP-CA model: the steep area with a slope over 25 degrees and the water space. The steep area is calculated from the DEM of China with a spatial resolution of 90 m, and the river area is abstracted from the national fundamental geographic information system (NFGIS, 1:4,000,000) of China.

# 3 The MVP-CA model

## *3.1 The model framework*

There are three modules in MVP-CA namely the macro module, the parcel generation module and the vector CA module (see Figure 4 for the flow chart). In the macro module, urban expansion rate is set for each city according to observed urban land expansion during 2007-2012 as the baseline scenario. The other two scenarios were set in accordance with famous spatial development strategies detailed in Section 3.2. In the parcel generation module, the parcels in 2012 (the base year of the model) were generated using AICP by Long and Liu (2013) for feeding the MVP-CA model as the base map (see Section 3.3). In the vector CA module, calibrated parameters abstracted from Beijing data and scenario settings by the macro module are fed to simulate future urban expansion during 2012-2017 (see Section 3.4).

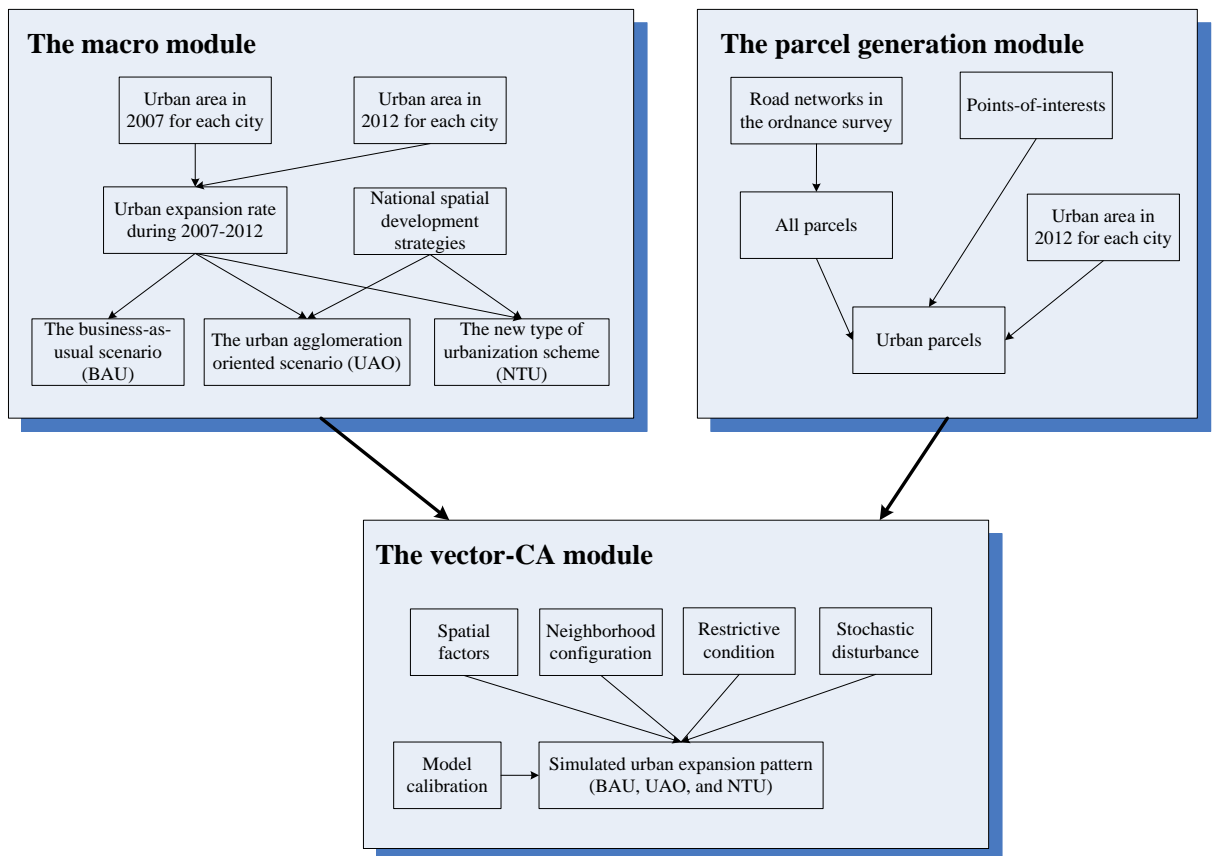

**Figure 4 The structure and flow diagram of MVP-CA**

## *3.2 The macro module*

It is not an easy task to precisely predict future urban expansion rates for Chinese cities. Alternatively, scenario analysis approach is adopted for simulating future urban expansion for all Chinese cities. Relevant studies on large scale in China show that policy regulations (especially the macro-regulation and regional development policies) and economic drivers (economic growth and population increase) are the leading causes of urban expansion (Liu et al, 2010). According to Deng et al (2010), China's urban land expands by 3% when the economy grows by 10%.

Considering the aforementioned points, three scenarios are fixed for the future development in this case study. The first one is the business-as-usual scenario (BAU) - a baseline scenario indicates that future urban expansion takes place based on the historic rules and current development tendency without any human adjustments. Urban land areas for each city in the designated year are qualitatively estimated depending on the growth rate calculated by utilizing the land statistics in 2007 and 2012.

The second scenario is the urban agglomeration oriented scheme (UAO), as indicated in the 11[th] Five-Year Plan and the 12[th] Five-Year Plan of National Economic and Social Development in China. In the next 5-10 years according to the plans, urban agglomerations in China will be developed as the main body of urbanization and the basic terrain unit in participating international competition and international division of labour (Wu et al, 2013). Considering the regional variance in China, the government has ratified and agreed to support more than 30 regional planning or development policies in urban agglomerations involving 23 provinces, autonomous regions and municipalities, which effectively promote the urban agglomeration towards a healthy and stable sustainable development in China. Based on above backgrounds and some related studies on urban agglomerations in China (Fang et al, 2013), 23 Urban Agglomerations (UAs) are considered as the second scenario analysis, which was shown in Figure 5. Those 355 cities in UAs are given a comparatively higher urban expansion rate of 5.0% while the other cities were given a relatively lower urban expansion rate of 4.0% in the next five years.

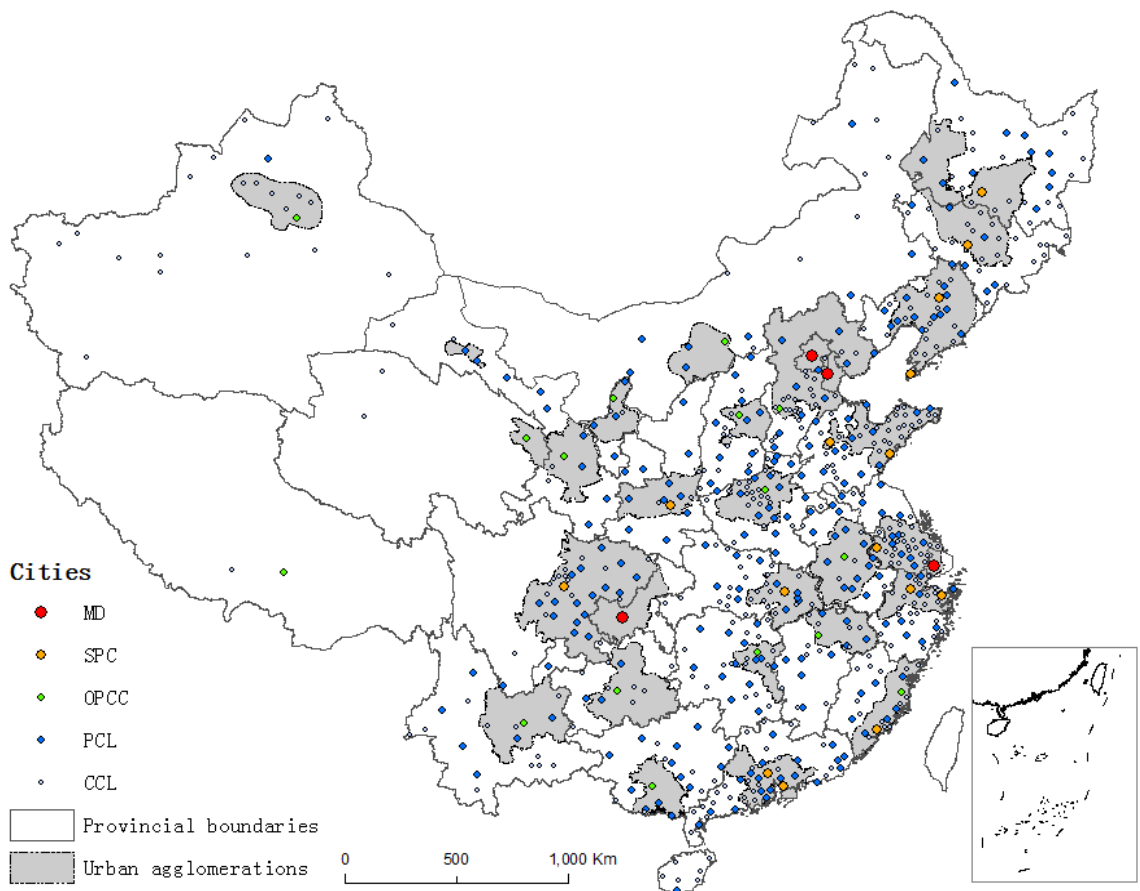

**Figure 5 The urban agglomerations in China**

The last scenario is the new type of urbanization scheme (NTU). The NTU is a macro future development roadmap that was announced by the central government in the end of 2013 in view of the development progress and emerging issues on China's urbanization process (Wu, 2013). It symbolizes the departure from land-centred urban development to people-oriented urbanization covering broad aspects. One of the key points is to foster the coordinated development between large, middle and small cities as well as small towns. In other words, small cities and towns shall be given priority in urbanization and urban development, while the big cities especially the megacities will be controlled step by step[4]. In this paper, we will set urban expansion rate of each city by its existing urban area. Considering city size classification in Chinese City Construction Statistics Yearbook, four types of cities are identified for the fowling five years as follows: super city whose urban land area above 400 km$^2$ in 2012 and with a urban expansion rate of 3.0% in; mega city whose urban land area between 200-400 km$^2$ in 2012 and with a urban expansion rate of 4.0%; big city whose urban land area between 100-200 km$^2$ in 2012 and with a urban expansion rate of 5.0%; medium-size and small city whose urban land area below 100 km$^2$ in 2012 and with a urban expansion rate of 6.0%.

### *3.3 The parcel generation module*

Urban parcels in 2012 are projected as the simulation start in MVP-CA. For this reason, IACP framework (Long and Liu, 2013) is adopted for inferring urban parcel distribution in 2012 of the whole China. In this manner, we use the surveyed roads for partitioning geographical space of China, and POIs for deducing density of each generated parcel. Then, vector cellular automata models help each city to select urban parcels out of all generated parcels, according to the total urban area available in MOHURD (2013). For more details on procedures, please check Long and Liu (2013).

### *3.4 The vector CA module*

According to the extracted urban expansion pattern during 1992-2008 in China estimated by Liu et al (2012) using DMSP/OLS nighttime light data, most of expanded urban land distributed on the periphery of existing urban land. This denotes that the distance to the city center and spatial adjacency between proposed area and existing urban land would significantly influence future urban expansion. Therefore, spatial constraints are taken into consider with vector-based constrained cellular automata (CA) for simulating urban expansion in each city[5] (Zhang and Long, 2013). Traditional CA consists of five components, namely a space represented as a regular grid composed of a collection of homogeneous cells, a set of possible cell states, and the transition rule which determines the evolution of the state of each cell based on statuses of its neighbouring cells and some external constraints at each time step (Batty et al, 1999; White and Engelen, 2000). In the proposed CA model, each parcel was regarded as a cell in CA, and the cell status was 0 (no expansion) or 1 (expanded from rural to urban). At the very beginning, all cells' status were set according to the 2012 parcels. Based on Feng et al (2011), the concept model of the proposed constrained CA was represented as $S_{ij}^{t+1} = f(S_{ij}^{t}, \Omega_{ij}^{t}, Con, N)$ ,

---

[4] The China Central Urbanization Working Conference was held during December 12-14, 2013 in Beijing. This conference bring forward the urbanization path of different kinds of cities in China: fully liberalize the settled restraints on small cities and towns; orderly open the settled limit on the medium-sized city, set reasonable settled conditions in big cities, and strictly control the scale of large urban population.

[5] Each city has its own constrained CA model for allocating urban parcels.

where $S_{ij}^{t}$ and $S_{ij}^{t+1}$ are the states of a cell *ij* time *t* and *t* + 1, respectively; *f* is the transition function; $\Omega_{ij}^{t}$ is the neighbourhood evaluation function; *Con* are the constraints on urban expansion; and *N* is the total number of cells. Every discrete time in CA is a year.

Specifically, the probability of a cell *ij* changing its state from non-urban to urban at time *t* can be represented as $P_{ij}^{t} = (P_{l})_{ij} \times (P_{\Omega})_{ij} \times con(\cdot) \times P_{r}$, where $(P_{l})_{ij}$ is a local potential a cell conversing from the non-urban to the urban; $(P_{\Omega})_{ij}$ is the state conversion potential of the cell within its neighbourhood; $con(\cdot)$ is the restrictive condition for urban development; and $P_{r}$ denotes the stochastic disturbance of any unknown errors (Feng et al, 2011).

The local potential $(P_{l})_{ij}$ can be determined through a set of factors using a logistic regression method (Wu, 2002):

$$(P_{l})_{ij} = \frac{1}{1+\exp[-(a_{0} + \sum_{k=1}^{m} a_{k}c_{k})]}$$

where $a_{0}$ is a constant, $c_{k}$ is a spatial variable, and $a_{k}$ is the estimated parameter/weight of $c_{k}$, and *m* is the amount of spatial variables. In this paper, we select four spatial factors, (1) the natural logarithm value of a parcel size (SIZE_LN), (2) the compactness of a parcel (COMPACT), calculated using Perimeter*Perimeter/Area, (3) the desired distance to city centres in km (CENTER), and (4) the POIs density calculated based on POIs (DENSITY). The density was standardized to range from 0 to 1 using the following equation: standardized density=log(raw)/log(max), where raw and max correspond to density of individual parcels and the nation-wide maximum density value.

Apart from other raster based constrained CA model[6], our model only contains two external spatial factors which are considered to be the most important factors driving a parcel's spatial expansion. Other factors like the desired distance to road networks could have been explained by the parcel itself already.

The state conversion potential of the cell within the neighbourhood can be defined as

$$(P_{W})_{ij} = \frac{\sum con(S_{ij}^{t} = urban)}{n}$$

where $con(S_{ij}^{t} = urban)$ represents the number of urban cells amongst the neighbourhood of cell *ij*, and *n* is the count of cells in the neighbourhood of cell *ij*. 500 m is dopted to identify the neighbouring relation between cells in this study.

Two layers - the steep area and various water bodies, are included as the restrictive condition. Urban expansion is forbidden in these areas. The constraints are expressed as $con(cell_{ij}^{t} = suitable)$ with a value of 0 or 1, where 1 indicates that there is no restriction on the parcel's development from rural to urban while 0 indicates that the parcel is forbidden from being developed as urban.

---

[6] The factors considered in the vector CA module are similar to Seto et al (2012), which uses slope, distance to roads, population density, and land cover as the primary drivers of land change. Since we use parcels as simulation units, and distance to roads is therefore not included in our module.

The stochastic disturbance $P_r$ in the model stands for any possible change of local policies and accidental errors. It is calculated using

$$P_r = 1 + (-\ln\gamma)^\beta.$$

where $\gamma$ is a random number ranging from 0 to 1, and $\beta$, ranging from 0 to 10, controls the effect of the stochastic factor.

By comparing the global probability $P_{ij}^t$ with a predefined threshold value $P_{thd}$ in the range of [0, 1], the model was then used to decide whether a non-urban cell can be converted to urban state at time $i+1$:

$$S_{ij}^{t+1} = \begin{cases} Urban \text{ for } P_{ij}^t \succ P_{thd} \\ NonUrban \text{ for } P_{ij}^t \leq P_{thd} \end{cases}.$$

## 3.5 Model validation

The model validation can be conducted module by module. The macro module as a rule-based module does not need to be validated. The baseline scenario is set based on the historical urban expansion data. Furthermore, the parcel generation module has already been validated in Long and Liu (2013). Hence, only the validation of the vector CA module has to be focused in this section from two perspectives. First, we compare the baseline scenario results in Beijing with the output of the BUDEM model (Long et al, 2009, Long et al, 2012[7]) which is a cell-based (500 m in square) urban expansion simulation model for Beijing. There are more spatial factors (market-oriented and institutional types) in the constrained CA based BUDEM, which has been successfully applied in various planning applications in Beijing. BUDEM would be calibrated using the same observed urban expansion for guaranteeing its comparability with the MVP-CA model in the city of Beijing. Second, in addition to the formal validation, the simulated urban expansion can also be validated by online browsers of the released parcel maps at CartoDB (an online WebGIS), in a manner of Wiki-map (Fritz et al, 2012). Browsers with local knowledge could engage and point out obvious simulation results and comment on the results. The comments are feedback to the authors for technical processing on the automated generated parcels.

# 4 Results

## 4.1 The parcels of all Chinese cities in 2012

The generated urban parcels in 2012 are shown in Figure 6. There are totally 761,152 urban parcels for all 654 Chinese cities and with a total land area 46,751 km² (the average urban parcel size is 6.1 hectares).

---
[7] There are slightly differences on BUDEM in the two papers. In this paper, we would follow Long et al (2012) in using BUDEM.

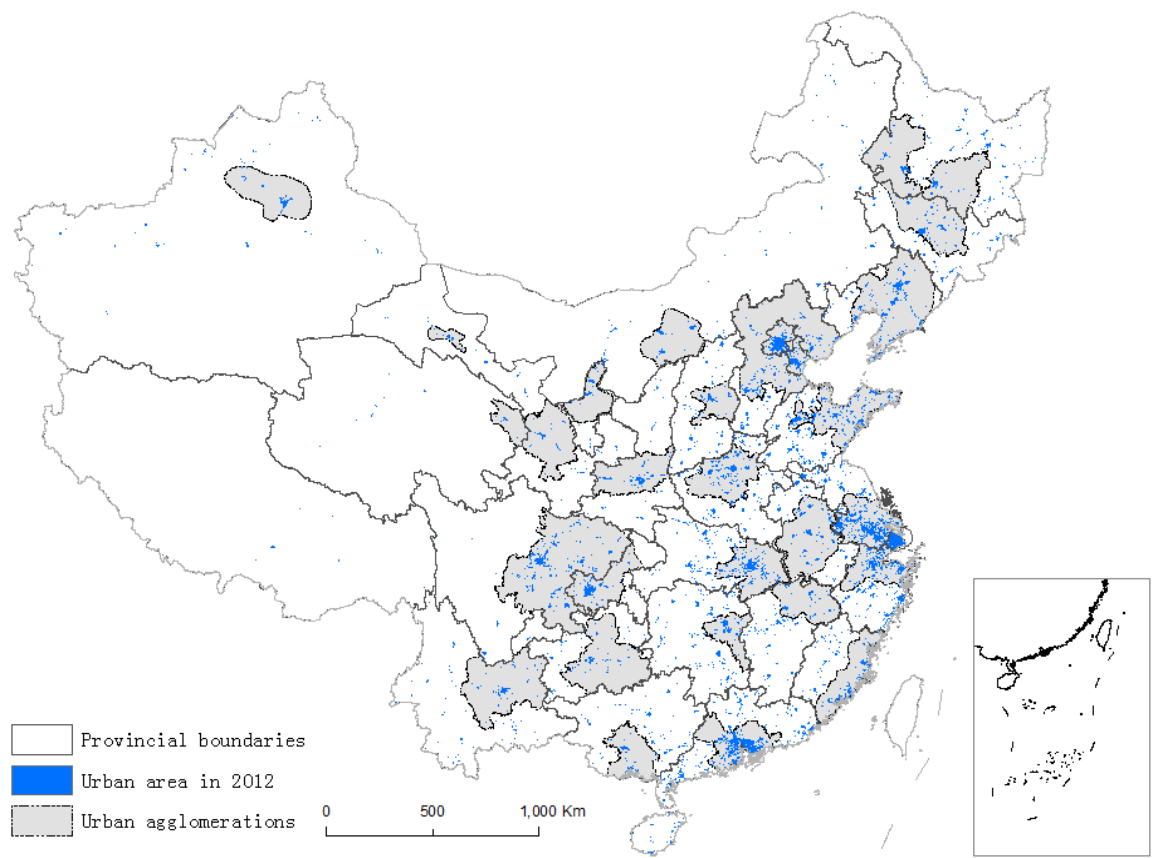

**Figure 6 Urban land areas of all Chinese cities**

## *4.2 Model calibration for the vector CA module*

Due to historical data availability, we do not have the whole country data for calibrating the vector CA module instead, we limit our model calibration within the city of Beijing (12,183 km$^2$, Yanqing and Miyun in the 16,410-km$^2$-Beijing Metropolitan Area not included), where both the parcel maps in 2007 and 2012 are obtained, in which urban land is available. Since the parcel geometries changed a lot during 2007-2012, only the intersected parcels are regarded as the samples (*N*=26877 including 5119 expanded parcels). The urban expansion is identified (expanded as 1, non-expanded as 0) and factor values are attached to all "intersected" parcels accordingly. Logistic regression is used for identifying parameters for the two spatial factors (density and city centre distance). The overall precision of logistic regression was 81.9%. The logistic regression results shown in Table 1 were applied in the MVP-CA model for all city regions[8]. The MVP-CA model was meanwhile used in the city of Beijing. An overall precision of 83.2% indicated the applicability of our CA model in replicating historical urban expansion in a city region.

---

[8] We admit the heterogeneity of weights in various city regions. We do not have existing parcels in other cities in writing this paper.

Table 1 Logistic regression results for the Beijing parcels

| Factor | Coefficient | S.E. | Wald | Sig. |
|---|---|---|---|---|
| SIZE_LN | -0.197 | 0.007 | 693.572 | 0.000 |
| COMPACT | 1.933 | 0.962 | 4.033 | 0.045 |
| CENTER | -.101 | 0.002 | 1891.809 | 0.000 |
| DENSITY | 2.230 | 0.110 | 407.554 | 0.000 |
| Constant | 2.224 | 0.082 | 739.440 | 0.000 |

## *4.3 Simulation results of the vector CA module*

Urban expansion patterns of China are simulated (Figure 7a-c). The BAU scenario presents a path-dependent urban expansion relying on the stable economic growth and land-use policies. Total urban land areas estimated by BAU are 62,835 km$^2$ in 2017, increased by 34.4% compared to 46,751 km$^2$ urban land in 2012. The overall spatial pattern of urban land in 2017 is similar to 2012. Some typical urban agglomerations in east will be developed the metropolitan interlocking regions (See Table 2). The simulated pattern in the UAO scenario indicates an urban agglomeration oriented scheme, and urban lands of those cities in UAs will expand more significantly no matter where they locate. Total urban land areas of UAO are 58,394 km$^2$ in 2017, increased by 24.9% compared to urban land in 2012, while 4,441 km$^2$ less than BAU. The last one is the NTU scenario, which considered the new type of urbanization scheme that was proposed by the central government of China in late 2013. Mega cities are strictly controlled for their urban land expansion that encroaches farmlands in the next five years. Although no notably difference in nationwide scope is portrayed in figure 6c, it is worth to note that medium and small -sized cities demonstrate a more rapid expansion than big ones. The total urban land areas of NTU are 58,930 km$^2$ in 2017, increased by 26.1% compared to urban land in 2012, while decreased 3,905 km$^2$ in comparison with BAU. Among all the three scenarios, the BAU scenario has the largest urban expansion ratio (34.4%) and may denote the upper limit of total urban land in China. Whereas spatial development strategies considered by the scenarios UAO and NTU have significant effects on curbing urban sprawl of Chinese cities by balancing both the sustainable supply of urban lands and demands of rapid urbanization.

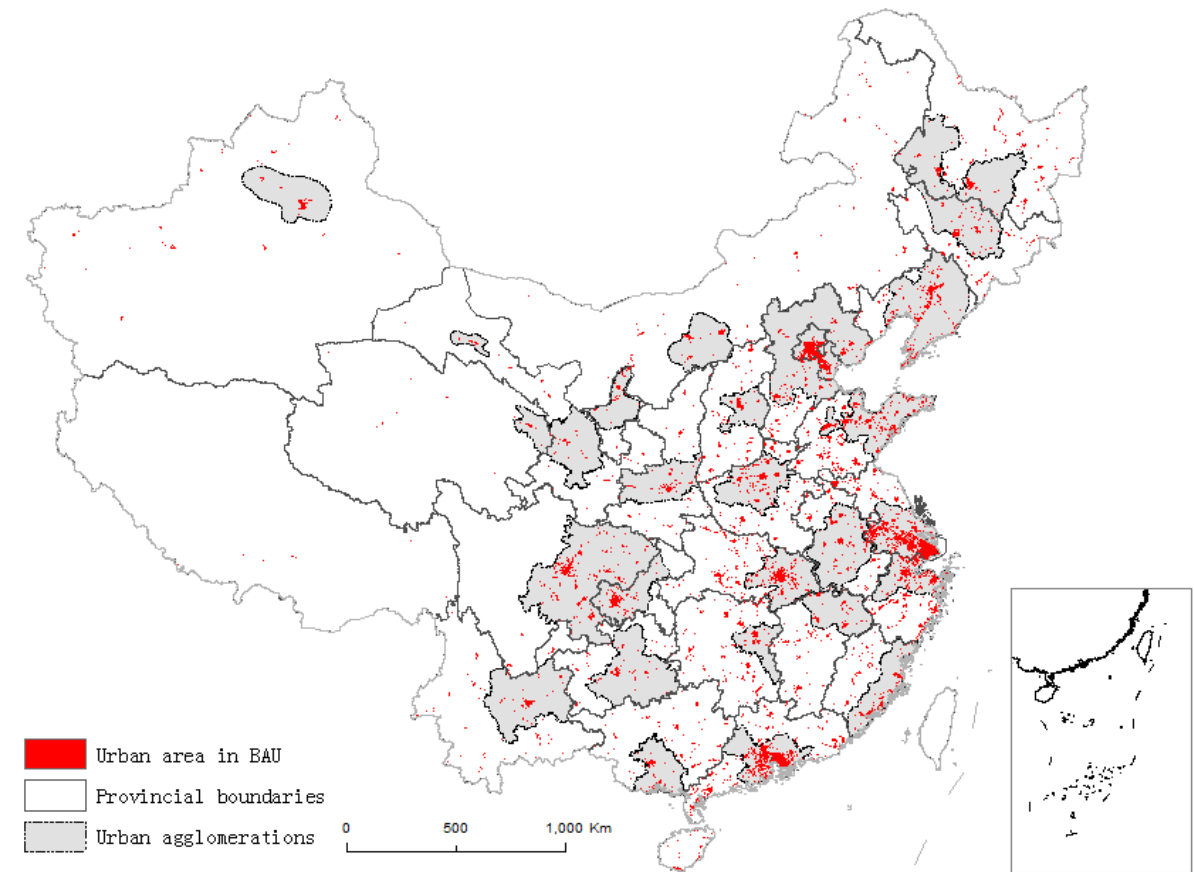

(a)

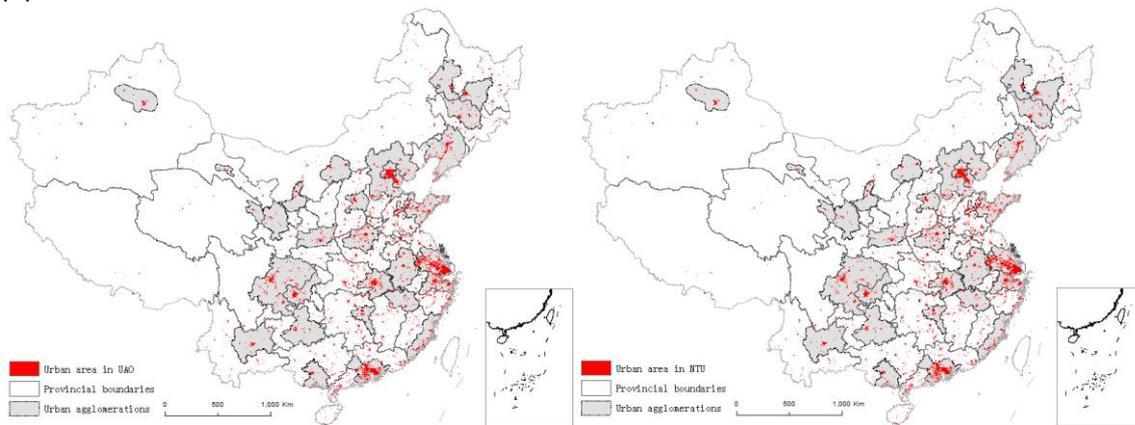

(b)                                (c)

**Figure 7 Urban area of all Chinese cities (a), and urban expansion patterns of the whole China for three scenarios (a: BAU, b: UAO, c: NTU)** [9]

Additionally, for gaining more details in simulation results of each scenario, urban land expansion patterns and total urban land areas of the three most important urban agglomerations: Beijing-Tianjin-Hebei (BTH), Yangtze River Delta (YRD) and Pearl River Delta (PRD) are listed in Table 2 for the further scrutiny. Without being exhaustive, several key features can be explored. First, according to the situations defined in UAO, the total quantity of urban land in UAO is the largest. By 2017, three urban agglomerations BTH, YRD, and PRD are projected to reach, respectively, 4,405 km$^2$, 9,144km$^2$, and 4,834 km$^2$ in urban land areas. Second, the three urban agglomerations in BAU do not have the largest total quantity of urban lands among the three scenarios, mainly because the urban expansion ratio is relatively lower in urban agglomerations before 2012 especially compared to those less developed regions who were experiencing a tremendous urbanization process. Third, the simulation results of NTU in the three urban agglomerations indicate a more sustainable pattern, and associate with the least total quantity of urban lands.

---

[9] For more vivid difference between scenarios of the whole China, please refer to Table 2 and online visualization.

**Table 2 Urban expansion patterns in typical urban agglomerations**

| UA | 1 BAU | 2 UAO | 3 NTU |
|---|---|---|---|
| BTH | 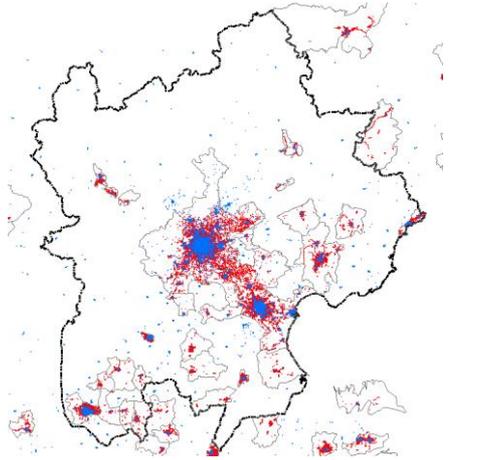 4,192 km² | 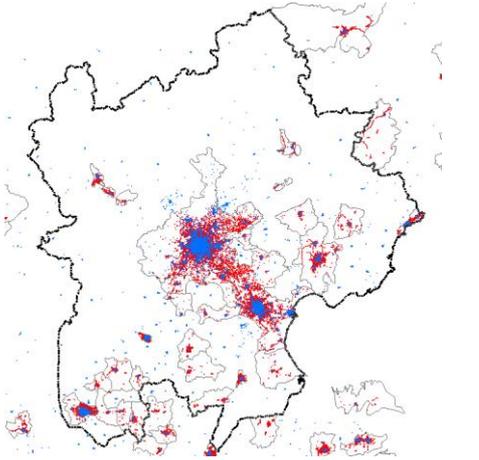 4,405 km² | 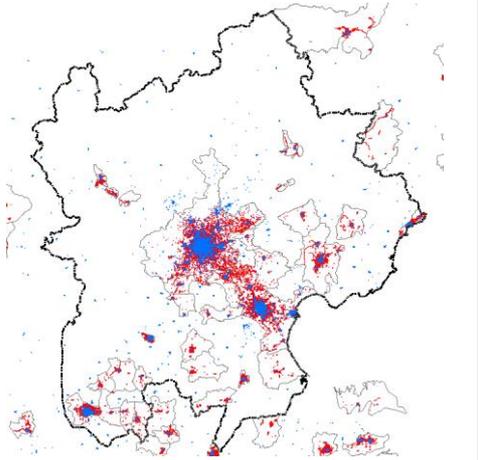 4,184 km² |
| YRD | 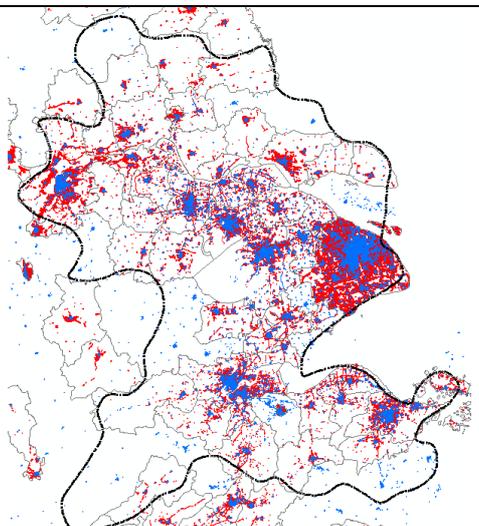 9,078 km² | 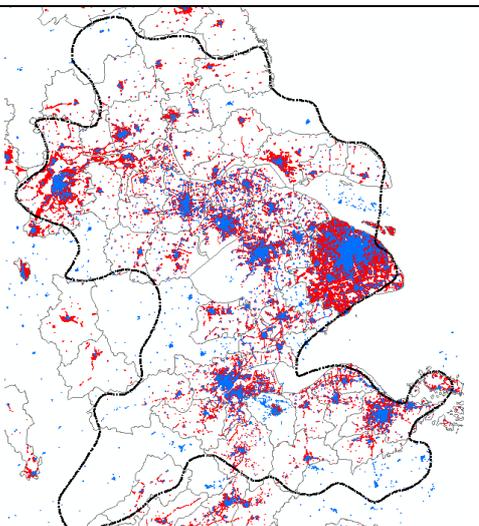 9,144 km² | 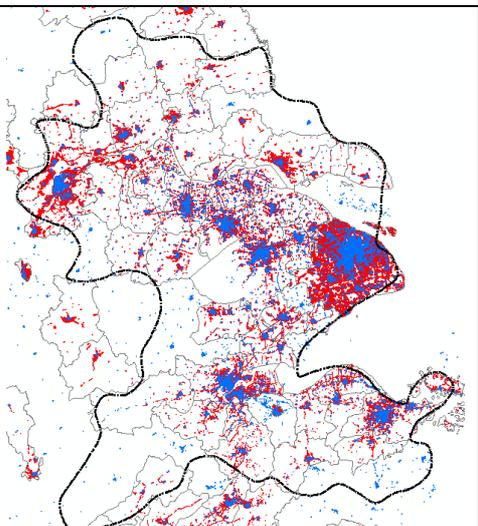 8,682 km² |
| PRD | 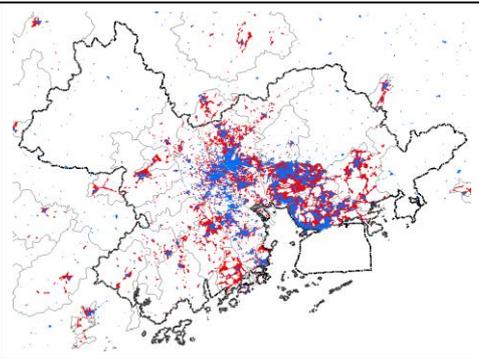 | 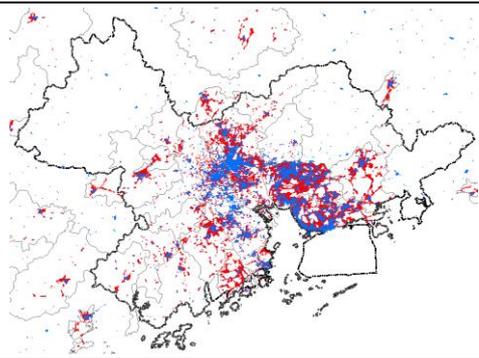 | 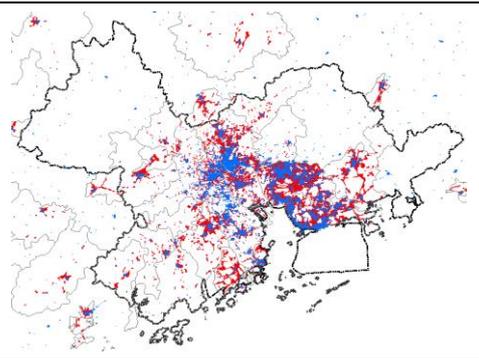 |
| | 4,660 km² | 4,834 km² | 4,519 km² |

Note that red denotes simulated urban expansion during 2012-2017 and blue denotes existing urban land in 2012.

Last but not least, we would release the simulation results of the three scenarios online on acceptance of this paper, and Table 3 shows urban expansion maps in the BAU scenario, captured from the future online visualization, for several typical cities.

**Table 3 The simulated results in the BAU scenario for typical cities**

| Zhengzhou | Beijing | Jinan | Shenyang |
|---|---|---|---|
| 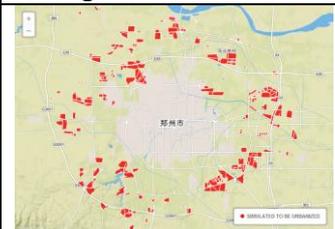 | 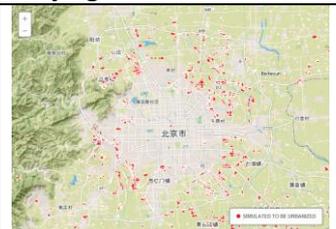 | 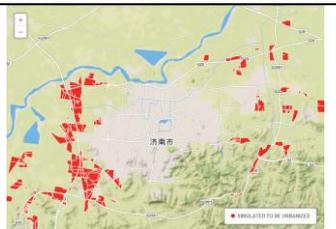 | 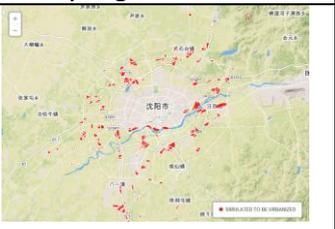 |
| Nanjing | Hangzhou | Shanghai | Wuhan |
| 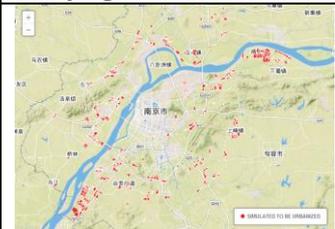 | 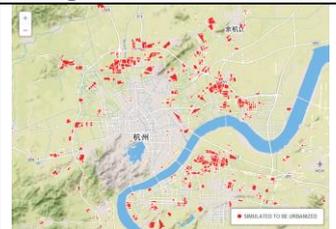 | 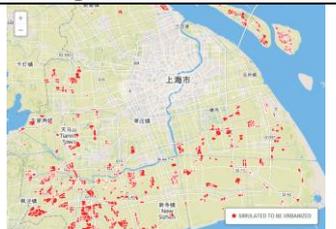 | 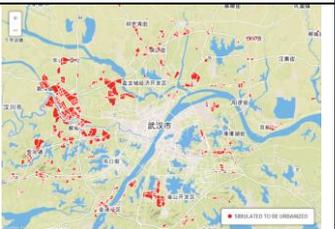 |
| Chengdu | Chongqing | Nanning | Guangzhou |
| 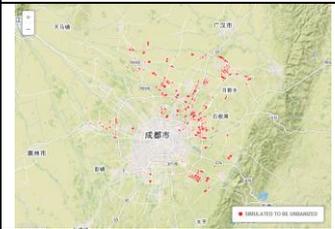 | 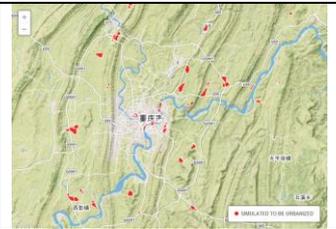 | 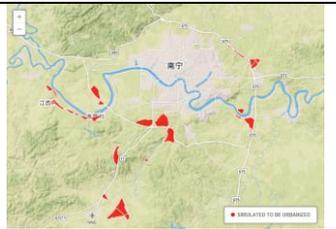 | 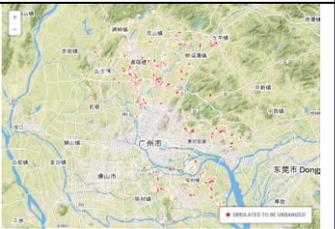 |

Note that red denotes simulated urban expansion during 2012-2017. The base map is the OpenStreetMap.

# 5 Discussion

## *5.1 The model validation*

In the first stage, the BAU scenario (limited to the city of Beijing, 12,183 km$^2$, the same with the model calibration of MVP-CA in Section 4.2) is validated with that by BUDEM. BUDEM was calibrated using the same urban parcels in 2007 and 2012 (also used in calibrating MVP-CA), in which parcels were converted into the 500m cells for feeding. The seven spatial factors remain the same with those in Long et al (2012). The overall precision in the logistic regression for parameter calibration of BUDEM was 96.1%, and calibrated parameters are shown in Table 4. The distance to the city centre (Tiananmen Square) is included in both models (CENTER in MVP-CA and l_tam in BUDEM). Cleary, its significance has Illustrated its influence on historical urban expansion in both model are both positive. That is to say, a place closer to the city centre tends to have a higher probability to be developed[10].

**Table 4 Logistic regression results for the Beijing parcels**

| Factor | Coefficient | S.E. | Wald | Sig. |
| --- | --- | --- | --- | --- |
| l_tam | 10.402 | 0.378 | 756.566 | 0.000 |
| l_city | 2.684 | 0.175 | 234.110 | 0.000 |
| l_town | -2.016 | 0.220 | 83.652 | 0.000 |
| l_road | 7.826 | 0.836 | 87.592 | 0.000 |
| g_conf | 0.535 | 0.089 | 35.774 | 0.000 |
| Constant | -11.832 | 0.801 | 218.329 | 0.000 |

Note: l_tam = the attractive potential of Tiananmen Square, l_city = the attractive potential of the closest new city centre, l_town = the attractive potential of the closest town centre, l_road = the attractiveness of the closest main roads, g_conf = whether or not a cell is forbidden for urban development. The other two factors in BUDEM, the attractive potential to the closest river l_river and whether or not a cell is farmland, are not significant in the regression.

In the following step, calibrated parameters of BUDEM was applied together with total urban area in 2017 and urban pattern in 2012 (converted into 500m cells) for Simulating urban expansion i Beijing. The total urban expansion area of Beijing was 174 km$^2$ (=1,619-1,445, the annual expansion ratio during 2007-2012 for Beijing was 2.3%) according to the settings in the BAU scenario of MVP-CA. Figure 8 shows the comparison results between the BAU scenario (expanded parcels) by MVP-CA and by BUDEM (expanded cells) in the city of Beijing. Expanded areas in both models exhibit similar patterns according to the visual judgement. When further examining the two expanded patterns in details, we recognized that he overlaid area shared by both patterns was 119 km$^2$ (68.4% of the total expansion) as shown in Figure 7. The overall precision in terms of the intersected results were acceptable considering the differences between two models. On the other hand, it is found that the simulated pattern for long term is not realistic due to developed large parcels when MVP-CA is adopted to predict for a longer time. This further proves that MVP-CA suits short-term urban expansion simulation in its current form, which does not include a parcel subdivision module yet.

---

[10] Note we use the desired distance for CENTER in MVP-CA and the influence potential in BUDEM, which makes the symbols are opposite.

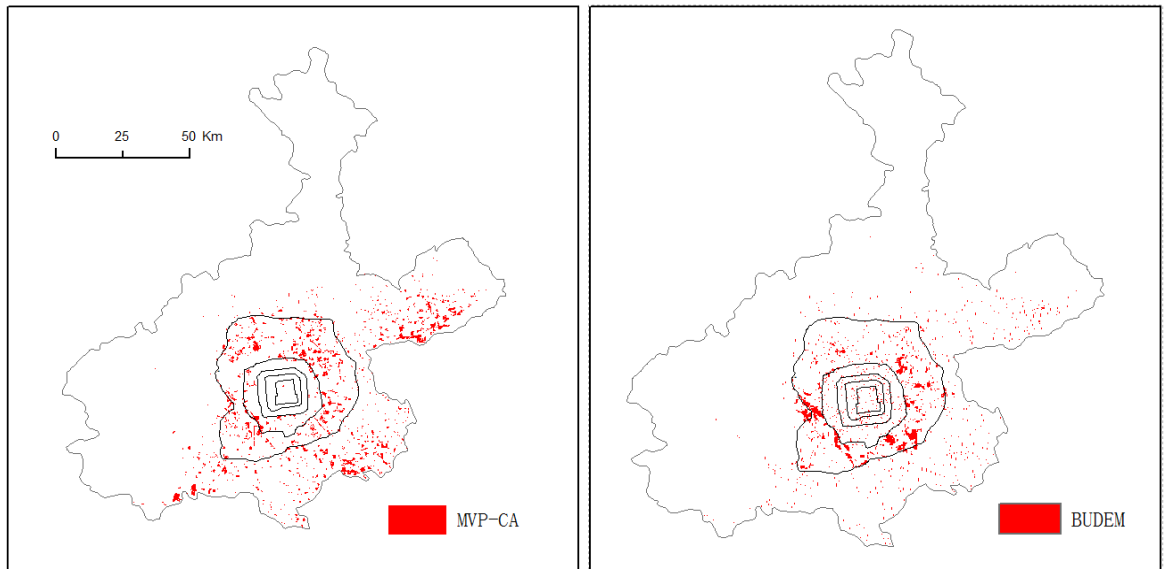

**Figure 8 Simulated urban expansion by MVP-CA and BUDEM in the city of Beijing**

For the online validation, our simulated-to-be-expanded parcels in the BAU were released online for user's comments via posting a message at Sina Weibo known as theChinese Twitter. The unique-ID was associated with each visualized parcel, thus enabling a browser pointing out a comment associate with the parcel. Generally, the simulated urban expansion by the parcel-based MVP-CA was attractive to browsers since its geometrical correspondence to readers' knowledge and local sense about the built environment compared with an abstracted square cell with no geography meanings (as revealed by Figure 7 and 8 as an example). In a period of three weeks after the results released, over 83 reposts and 76 comments[11] were received. Most of comments were positive on our study and pointed out such efforts were in need by developers, planners and decision makers. The President of China Planning Association commented that "every city should have its own spatial model to facilitate more scientific decision making in the process of urban development, and the research falls into the pool of planning section of Chinese Dream". Two comments said several simulated-to-expanded parcels were not plausible to develop due to strict local policies on development. Four comments said the region highlighted by local governments by means of spatial plans or development policies should be taken into account in the model. For an instance of Taiyuan, as mentioned by a browser whose hometown is Taiyuan, most of new expansions are simulated to distribute in the northern part. According to the local ambition emphasized by the Taiyuan government, however, the northern part of the city has been paid more attention by the government. The simulated urban expansion would be further enhanced with more comments received from participators with local experience and knowledge to the physical environment.

## *5.2 Potential bias and next steps*

---

[11] Not like Twitter, Sina Weibo also allows users to comment on a post, rather than only reposting it.

Although admitting the merits of this study, there are several limitations in the present study. **First**, the parcel generation process would be enhanced via incorporating with other existing land cover dataset for generating more accurate existing urban parcels. We have a plan to include the global land cover data FROM-GLC-agg by Yu et al (2013) as well as the urban land inferred from DMSP/OLS by Yang et al (2013) for assisting urban parcel generation. **Second**, there are some large parcels developed in the simulation results, especially in small cities, which is not quite realistic in the real world. Techniques for parcel subdivision would be an alternative solution for generating more practical urban parcels in China (Aliaga et al, 2008; Wickramasuriya et al, 2013). **Third**, the model should be calibrated using national datasets, rather than being limited to Beijing in this study thereby enhancing the model simulation precision.

With the aforementioned potential biases in mind, the future study would focus on the following two aspects. First, a spatial equilibrium module considering the provincial level input-output would replace the current "the macro module" in the near future. The integration of equilibrium mechanism with the dynamic CA model enables linking the inter-provincial, even inter-city, simulation in the macro level and urban expansion simulation in the locally parcel level. Second, as commented by browsers, local development policies and spatial plans of various cities could be added as a factor in the vector CA module to reflect local development policies.

### *5.3 Potential applications*

The established national-scale urban expansion simulation model for the whole China, together with three simulated scenarios, could be applied in but not limited to the following aspects. **First**, national spatial development strategies, reflected in the form of variation of each city's urban expansion rate, could be visualized at a fine scale using MVP-CA, via linking macro policies to local developments. Although we only simulated three scenarios reflecting macro strategies in this paper, other spatial policies in local cities are also possible to be evaluated via adjusting the parameters of spatial factors in MVP-CA. **Second,** parcel-level simulation results can directly aware local stakeholders' places of interest on future developments, which is not easy by simulation results by other spatial expansion models with large simulation units. We have shared the results online thus helping promoting this application while the feedbacks could be absorbed in model adjustments. **Third**, the simulation results provide it possible to conduct urban expansion impact analysis, e.g. environment, ecology and social impacts. With the release of the simulated urban expansion scenarios, we expect it would attract more researchers to apply the simulated patterns for addressing their impacts in various avenues.

## 6 Concluding remarks

In this paper, a mega-vector-parcels cellular automata (MVP-CA) is developed for simulating urban expansion of all 654 Chinese cities. Three modules, the macro module, the parcel generation module, and the vector CA module, were equipped in MVP-CA. The macro module was responsible for setting urban expansion rate in the next five years for each city, taking into account both historical urban expansion rate and national spatial development strategies. The parcel generation module was used for identifying existing urban parcels in 2012 using the framework of AICP (automatic identification and characterization of parcels) proposed by Long and Liu (2013). The vector CA module was applied for simulating urban expansion in 2012-2017, and calibrated using the urban expansion data in Beijing. Three urban expansion scenarios, baseline (BAU), urban agglomeration (UAO), and new urbanization development (NTO), have been simulated for 2012-2017 by MVP-CA, respectively. We validated the simulation results by two solutions, the first was to compare the baseline scenario of Beijing with that by a raster CA model BUDEM we developed before, and the second was to validate the results via a wiki-manner.

As the first large-scale urban expansion model in the fine-scale for the whole China, our contributions mainly lie in the following aspects. **First**, a vector-based cellular automata model was introduced for simulating urban expansion in a super large geographical scale in the parcel level, which is rare in existing applied urban models. **Second**, we proposed a solution for linking national spatial development strategies with urban expansion via reflecting as the urban expansion speed of each city. This enables simulating macro policies in a fine-scale through the channel of MVP-CA. **Last**, we simulated the near-future urban area for all Chinese cities, which, together with existing urban area, would be shared online as an important data infrastructure for both practitioners and researchers.

The computation time has been a bottleneck for vector CA model. In this paper, the key computation load attributes to calculating neighbouring parcels of each parcel. It took around 3 days by a standalone computer, for 851,054 parcels in 654 cities, which is facilitated by ArcGIS using Python. This process is automatic and only needs to be calculated for one time. Once calculated, we do not need touch it during all simulation process, thus the computation load problem could be therewith solved. It is also recognise that separating the neighbourhood calculation by city would decrease the computation time consumed significantly (from previous 12 days for all parcels simultaneously loaded into memory to the final 3 days). After the neighbourhood preparation, it took around 20 hours for simulating urban expansion for each scenario, which is acceptable for a national scale urban expansion model in the parcel level.